\documentclass[submission,copyright,creativecommons,noderivs]{eptcs}
\providecommand{\event}{ACCAT 2012} 
\usepackage{breakurl}     

\title{Decorated proofs for computational effects: States}
\date{January 27., 2011}
\author{
  Jean-Guillaume Dumas 
\thanks{This work is partly funded by the project HPAC 
of the French Agence Nationale de la Recherche (ANR 11 BS02 013).} 
\quad 
  Dominique Duval
\thanks{This work is partly funded by the project CLIMT 
of the French Agence Nationale de la Recherche (ANR 11 BS02 016).} 
 \quad 
  Laurent Fousse \quad 
  Jean-Claude Reynaud 
  \institute{LJK, Universit\'e de Grenoble, France.} \\
  \email{\{Jean-Guillaume.Dumas, Dominique.Duval,
  Laurent.Fousse, Jean-Claude Reynaud\}@imag.fr}
}	

\def\titlerunning{Decorated proofs for computational effects: States}
\def\authorrunning{J.-G. Dumas, D. Duval, L. Fousse \& J.-C. Reynaud}

\usepackage{amsmath}
\usepackage{amsthm} 
\usepackage{amssymb}
\usepackage{enumerate}
\usepackage{hyperref}
\usepackage{nonfloat}
\usepackage{rotating}
\usepackage{url}
\usepackage{xspace}
\usepackage{bussproofs}
\usepackage{mathbbol}
\usepackage[all]{xy}
\SelectTips{cm}{}

\theoremstyle{plain}
\newtheorem{thm}{Theorem}[section]
\newtheorem{prop}[thm]{Proposition}
\theoremstyle{definition}

\newtheorem{rem}[thm]{Remark}

\newcommand{\pure}{{(0)}} 
\newcommand{\eqs}{\equiv} 
\newcommand{\eqw}{\sim} 
\newcommand{\acc}{{(1)}} 
\newcommand{\modi}{{(2)}} 
\newcommand{\Loc}{\mathit{Loc}} 
\newcommand{\Val}{\mathit{Val}} 
\newcommand{\St}{\mathit{St}} 
\newcommand{\sfl}{\mathit{lookup}} 
\newcommand{\sfu}{\mathit{update}} 
\newcommand{\perm}{\mathit{perm}} 

\newcommand{\unit}{\mathbb{1}}
\newcommand{\tuple}[1]{\langle #1 \rangle}
\newcommand{\tu}{\tuple{\,}}

\newcommand{\Ss}{\Sigma} 
\newcommand{\Tt}{\Theta} 
\newcommand{\Ff}{\Phi} 
\newcommand{\Tbank}{\mathit{Bank}} 
\newcommand{\Tstate}{\mathit{State}} 
\newcommand{\Tset}{\mathit{Set}} 

\newcommand{\deco}{\mathrm{deco}} 
\newcommand{\app}{\mathrm{app}} 
\newcommand{\expl}{\mathrm{expl}} 

\newcommand{\bZ}{\mathbb{Z}} 
\newcommand{\bU}{\{\star\}} 

\newcommand{\catT}{\mathbf{T}} 

\newcommand{\deno}[1]{[[#1]]}
\newcommand{\hsp}{\null\hspace{10pt}}
\newcommand{\cpp}{\texttt{C++}\xspace}
\newcommand{\id}{\mathit{id}}
\newcommand{\logL}{\mathcal{L}}

\newcommand{\rn}[1]{\;\textrm{(#1)}\;} 

\newcommand{\rncomp}{\rn{comp}}
\newcommand{\rnid}{\rn{id}}
\newcommand{\rnassoc}{\rn{assoc}}
\newcommand{\rnidsrc}{\rn{id-src}}
\newcommand{\rnidtgt}{\rn{id-tgt}}

\newcommand{\rnsym}{\rn{$\equiv$-sym}}
\newcommand{\rntrans}{\rn{$\equiv$-trans}}

\newcommand{\rntuple}{\rn{tuple}}
\newcommand{\rnproj}{\rn{tuple-proj-$i$}}
\newcommand{\rntupleun}{\rn{tuple-unique}}
\newcommand{\rnfinal}{\rn{final}}
\newcommand{\rnfinalun}{\rn{final-unique}}

\newcommand{\rnsrefl}{\rn{$\eqs$-refl}}
\newcommand{\rnssym}{\rn{$\eqs$-sym}}
\newcommand{\rnstrans}{\rn{$\eqs$-trans}}
\newcommand{\rnssubs}{\rn{$\eqs$-subs}}
\newcommand{\rnsrepl}{\rn{$\eqs$-repl}}
\newcommand{\rnpa}{\rn{0-to-1}}
\newcommand{\rnpcomp}{\rn{0-comp}}
\newcommand{\rnacomp}{\rn{1-comp}}
\newcommand{\rnpid}{\rn{0-id}}
\newcommand{\rnwrefl}{\rn{$\eqw$-refl}}
\newcommand{\rnwsym}{\rn{$\eqw$-sym}}
\newcommand{\rnwtrans}{\rn{$\eqw$-trans}}
\newcommand{\rnwsubs}{\rn{$\eqw$-subs}}
\newcommand{\rnwrepl}{\rn{0-$\eqw$-repl}}
\newcommand{\rnsw}{\rn{$\eqs$-to-$\eqw$}}
\newcommand{\rnws}{\rn{1-$\eqw$-to-$\eqs$}}
\newcommand{\rnpfinal}{\rn{0-final}}  
\newcommand{\rnwfinalun}{\rn{$\eqw$-final-unique}}
\newcommand{\rnatuple}{\rn{obs-tuple}} 
\newcommand{\rnaproj}{\rn{obs-tuple-proj-$i$}} 
\newcommand{\rnstupleun}{\rn{obs-tuple-unique}} 

\newcommand{\rnpprod}{\rn{0-prod}}
\newcommand{\rnpprojone}{\rn{0-proj-1}}
\newcommand{\rnpprojtwo}{\rn{0-proj-2}}
\newcommand{\rnpprodun}{\rn{0-prod-unique}}
\newcommand{\rnlprod}{\rn{left-prod}}
\newcommand{\rnlprojone}{\rn{left-proj-1}}
\newcommand{\rnlprojtwo}{\rn{left-proj-2}}
\newcommand{\rnlprodun}{\rn{left-prod-unique}}
\newcommand{\rnrprod}{\rn{right-prod}}
\newcommand{\rnrprojone}{\rn{right-proj-1}}

\begin{document}

\maketitle

\begin{abstract}
\textbf{Abstract.} 
The syntax of an imperative language  
does not mention explicitly the state, 
while its denotational semantics has to
mention it. 
In this paper we show that the equational proofs about  
an imperative language may hide the state, in the 
same way as the syntax does. 
\end{abstract}

\section*{Introduction}

The evolution of the state of the memory 
in an imperative program is a computational effect: 
the state is never mentioned as an argument or a result 
of a command, whereas in general it is used and
modified during the execution of commands. 
Thus, the syntax of an imperative language  
does not mention explicitly the state, 
while its denotational semantics has to mention it. 
This means that the state is \emph{encapsulated}: 
its interface, which is made of the functions for looking up 
and updating the values of the locations, 
is separated from its implementation; 
the state cannot be accessed in any other way than 
through his interface. 
In this paper we show that equational proofs in 
an imperative language may also encapsulate the state: 
proofs can be performed without any knowledge of 
the implementation of the state. 
We will see that a naive approach (called ``apparent'') 
cannot deal with the updating of states,
while this becomes possible with 
a slightly more sophisticated approach (called ``decorated''). 
This is expressed in an algebraic framework relying on category theory. 
To our knowledge, the first categorical treatment of 
computational effects, using monads, 
is due to Moggi \cite{Mo91}. 
The examples proposed by Moggi include 
the side-effects monad $T(A) = (A\times \St)^\St$ 
where $\St$ is the set of states. 
Later on, Plotkin and Power 
used Lawvere theories for dealing with 
the operations and equations related to computational effects.   
The Lawvere theory for the side-effects monad  
involves seven equations \cite{PP02}. 
In Section~\ref{sec:motivations} we describe the intended 
denotational semantics of states. 
Then in Section~\ref{sec:states} we
introduce three variants of the equational logic for formalizing the 
computational effects due to the states:
the \emph{apparent}, \emph{decorated} an \emph{explicit} logics. 
This approach is illustrated 
in Section~\ref{sec:proofs} by proving some of the equations 
from \cite{PP02}, using rules which do not mention any 
type of states.

\section{Motivations} 
\label{sec:motivations}

This section is made of three independent parts. 
Section~\ref{subsec:states} is devoted to the semantics of states, 
an example is presented in Section~\ref{subsec:bank},
and our logical framework is described in Section~\ref{subsec:dialog}.

\subsection{Semantics of states}
\label{subsec:states}

This section deals with the denotational semantics of states, 
by providing a set-valued interpretation of the 
\emph{lookup} and \emph{update} operations.
Let $\St$ denote the set of \emph{states}. 
Let $\Loc$ denote the set of \emph{locations}
(also called \emph{variables} or \emph{identifiers}). 
For each location $i$, 
let $\Val_i$ denote the set of possible \emph{values} for $i$. 
For each location $i$ there is a \emph{lookup} function 
for reading the value of location $i$ in the given state,
without modifying this state: 
this corresponds to a function $\sfl_{i,1}:\St\to\Val_i$
or equivalently to a function $\sfl_i:\St\to\Val_i\times\St$ 
such that $\sfl_i(s)=\tuple{\sfl_{i,1}(s),s}$ for each state $s$.
In addition, for each location $i$ there is
an \emph{update} function $\sfu_i:\Val_i\times\St\to\St$ for  
setting the value of location~$i$ to the given value,
without modifying the values of the other locations in the given state. 
This is summarized as follows, for each $i\in\Loc\,$:
a set $\Val_i$, 
two functions $\sfl_{i,1}:\St\to\Val_i$ and $\sfu_i:\Val_i\times\St\to\St$, 
and equations $ (1) $:
\begin{itemize}
\item[(1.1)] 
$\forall\,a\in \Val_i \,,\; \forall\,s\in \St\,,\;  
     \sfl_{i,1}(\sfu_i(a,s)) = a \;,  $
\item[(1.2)] 
$\forall\,a\in \Val_i \,,\; \forall\,s\in \St \,,\;   
     \sfl_{j,1}(\sfu_i(a,s)) = \sfl_{j,1}(s) \;
     \mbox{ for every } j \in \Loc,\; j\ne i \;. $
\end{itemize}
The state can be observed thanks to the lookup functions.
We may consider the tuple 
$\tuple{\sfl_{i,1}}_{i\in\Loc}:\St\to\prod_{i\in\Loc}\Val_i$. 
If this function is an isomorphism, then Equations~(1)  
provide a definition of the update functions.
In \cite{PP02} an equational presentation of states is given, 
with seven equations: in Remark~\ref{rem:states-eq} 
these equations are expressed according to \cite{Me10}
and they are translated in our framework. 
We use the notations $l_i=\sfl_i:\St\to\Val_i\times\St$, 
$l_{i,1}=\sfl_{i,1}:\St\to\Val_i$ and $u_i=\sfu_i:\Val_i\times\St\to\St$, 
and in addition $\id_i:\Val_i\to\Val_i$ and $q_i:\Val_i\times\St\to\St$ 
respectively denote the identity of $\Val_i$ and the projection,
while 
$\perm_{i,j}:\Val_j\times\Val_i\times\St\to\Val_i\times\Val_j\times\St$ 
permutes its first and second arguments. 

\begin{rem}
\label{rem:states-eq} 
The equations in \cite{PP02} can be expressed as the following Equations (2):
\begin{itemize}
\item[(2.1)] 
Annihilation lookup-update.
\textit{Reading the value of
a location $i$ and then updating the location $i$ with the
obtained value is just like doing nothing.}
\\ $\hsp$ $ \forall\, i\in\Loc ,\;\forall\, s\in\St ,\;\; 
  u_i(l_i(s)) = s 
  \in\St $
\item[(2.2)] 
Interaction lookup-lookup.
\textit{Reading twice the same
location loc is the same as reading it once.}
\\ $\hsp$ $ \forall\, i\in\Loc ,\;\forall\, s\in\St ,\;\;
  l_i(q_i(l_i(s))) = l_i(s) 
  \in\Val_i\times\St $
\item[(2.3)] 
Interaction update-update. 
\textit{Storing a value $a$ and then a value $a'$ 
at the same location~$i$ is just like storing
the value $a'$ in the location.} 
\\ $\hsp$ $ \forall\, i\in\Loc ,\;\forall\, s\in\St ,\;\forall\, a,a'\in\Val_i
  ,\;\; u_i(a',u_i(a,s)) = u_i(a',s) 
  \in\St $  
\item[(2.4)] 
Interaction update-lookup. 
\textit{When one stores a
value $a$ in a location $i$ and then reads the location $i$,
one gets the value $a$.} 
\\ $\hsp$ $ \forall\, i\in\Loc ,\;\forall\, s\in\St ,\;\forall\, a\in\Val_i 
  ,\;\; l_{i,1}(u_i(a,s)) = a 
  \in\Val_i $
\item[(2.5)] 
Commutation lookup-lookup.
\textit{The order of reading
two different locations $i$ and $j$ does not matter.} 
\\ $\hsp$ $ \forall\, i\ne j\in\Loc ,\;\forall\, s\in\St ,\;\; 
  (\id_i\times l_j)(l_i(s)) = \perm_{i,j}((\id_j\times l_i)(l_j(s))) 
  \in\Val_i\times\Val_j\times\St $ 
\item[(2.6)] 
Commutation update-update.
\textit{The order of storing
in two different locations $i$ and $j$ does not matter.}  
\\ $\hsp$ $ \forall\, i\ne j\in\Loc ,\;\forall\, s\in\St ,\;
  \forall\, a\in\Val_i ,\;\forall\, b\in\Val_j ,\;\; 
  u_j(b,u_i(a,s)) = u_i(a,u_j(b,s)) 
  \in\St $ 
\item[(2.7)] 
Commutation update-lookup.
\textit{The order of storing
in a location $i$ and reading in another location~$j$ does not
matter.} 
\\ $\hsp$ $ \forall\, i\ne j\in\Loc ,\;\forall\, s\in\St 
  ,\;\forall\, a\in\Val_i ,\;\; 
  l_j(u_i(a,s)) = (\id_j\times u_i)(\perm_{j,i}(a,l_j(s)))
  \in \Val_j\times\St $
\end{itemize} 
\end{rem}

\begin{prop}
\label{prop:states-eq}  
Let us assume that $\tuple{l_{i,1}}_{i\in\Loc}:\St\to\prod_{i\in\Loc}\Val_i$ 
is invertible. 
Then Equations~(1) are equivalent to Equations~(2).
\end{prop}

\proof 
It may be observed that (2.4) is exactly (1.1).
In addition, (2.7) is equivalent to (1.2)~:  
indeed, (2.7) is equivalent to the conjunction of 
its projection on $\Val_j$ and its projection on $\St$;
the first one is $l_{j,1}(u_i(a,s))=l_{j,1}(s)$,
which is (1.2), and the second one is $u_i(a,s) = u_i(a,s)$. 
Equations (2.2) and (2.5) follow from $q_i(l_i(s))= s$. 
For the remaining equations (2.1), (2.3) and (2.6),  
which return states, it is easy to check that for each location $k$, 
by applying $l_k$ to both members and using equation~(1.1) or (1.2)
according to $k$, we get the same value in $\Val_k$ for both hand-sides.
Then equations (2.1), (2.3) and (2.6) follow from 
the fact that $\tuple{l_{i,1}}_{i\in\Loc}:\St\to\prod_{i\in\Loc}\Val_i$ 
is invertible. 
\qed 

Proposition~\ref{prop:states-eq} will be revisited in Section~\ref{sec:proofs},
where it will be proved that equations (1) imply 
equations (2) without ever mentioning explicitly the state
in the proof.

\subsection{Computational effects: an example} 
\label{subsec:bank}

In an informal way, 
we consider that a computational effect occurs when there is 
an apparent mismatch, i.e., some lack of soundness, 
between the syntax and the denotational semantics of a language. 
For instance in an object-oriented language, the state of an object  
does not appear explicitly as an argument nor as a result 
of any of its methods. 
In this section, as a toy example, we build a class \texttt{BankAccount} 
for managing (very simple!) bank accounts. 
We use the types \texttt{int} and \texttt{void}, 
and we assume that \texttt{int} is interpreted by the set of integers $\bZ$ 
and \texttt{void} by a singleton $\bU$. 
In the class \texttt{BankAccount}, there is a method 
\texttt{balance()} which returns the current balance of the account
and a method \texttt{deposit(x)} for the deposit 
of \texttt{x} Euros on the account. 
The \texttt{deposit} method is a \emph{modifier},
which means that it can use and modify the state of the current account.
The \texttt{balance} method 
is an \emph{inspector}, or an \emph{accessor}, which means that 
it can use the state of the current account but it is not allowed to modify 
this state.
In the object-oriented language \cpp,
a method is called a \emph{member function};
by default a member function is a modifier, 
when it is an accessor it is called a \emph{constant member function}
and the keyword \texttt{const} is used.
So, the \cpp syntax for declaring the member functions
of the class \texttt{BankAccount} looks like:
\begin{center}
\begin{tabular}{l}
  \texttt{int} \texttt{balance} (\,)  \texttt{const} ; \\
  \texttt{void} \texttt{deposit} (\texttt{int}) ; \\
\end{tabular}
\end{center}
\begin{itemize}
\item 
Forgetting the keyword \texttt{const}, 
this piece of \cpp syntax can be translated as a signature $\Tbank_\app$,
which we call the \emph{apparent signature}
(we use the word ``apparent'' in the sense of ``seeming''
i.e., ``appearing as such but not necessarily so'').  
$$ \Tbank_\app : \begin{cases}
  \texttt{balance}:\texttt{void}\to\texttt{int} \cr
   \texttt{deposit}:\texttt{int}\to\texttt{void} 
 \end{cases} $$
In a model (or algebra) of the signature $\Tbank_\app$, 
the operations would be interpreted as functions:
$$
  \begin{cases}
  \deno{\texttt{balance}}:\bU\to\bZ \cr
  \deno{\texttt{deposit}}:\bZ\to\bU 
 \end{cases} 
$$
which clearly is not the intended interpretation.
\item 
In order to get the right semantics, 
we may use another signature $\Tbank_\expl$, 
which we call the \emph{explicit signature},
with a new symbol \texttt{state} for the ``type of states'':
$$ \Tbank_\expl : \begin{cases}
  \texttt{balance}:\texttt{state}\to\texttt{int} \cr
   \texttt{deposit}:\texttt{int}\times\texttt{state}\to\texttt{state} 
 \end{cases} $$
The intended interpretation is a model of the explicit signature $\Tbank_\expl$, 
with $\St$ denoting the set of states of a bank account:
$$ \begin{cases}
  \deno{\texttt{balance}}:\St\to\bZ \cr
  \deno{\texttt{deposit}}:\bZ\times\St\to\St 
 \end{cases} $$
\end{itemize}
So far, in this example, we have considered two different signatures.
On the one hand, the apparent signature $\Tbank_\app$
is simple and quite close to the \cpp code, but 
the intended semantics is not a model of  $\Tbank_\app$.
On the other hand, the semantics is a model of the explicit 
signature $\Tbank_\expl$, 
but $\Tbank_\expl$ is far from the \cpp syntax: actually, the very nature of 
the object-oriented language is lost by introducing a ``type of states''. 
Let us now define a \emph{decorated signature} $\Tbank_\deco$,
which is still closer to the \cpp code than the apparent signature 
and which has a model corresponding to the intended semantics.
The decorated signature is not exactly a signature in the classical sense, 
because there is a classification of its operations.
This classification is provided by superscripts called \emph{decorations}: 
the decorations \texttt{(1)} and \texttt{(2)} correspond respectively 
to the object-oriented notions of \emph{accessor} and \emph{modifier}.  
$$ \Tbank_\deco :   \begin{cases}
  \texttt{balance}^\texttt{(1)}:\texttt{void}\to\texttt{int} \cr
  \texttt{deposit}^\texttt{(2)}:\texttt{int}\to\texttt{void} 
  \end{cases} $$
The decorated signature is similar to the \cpp code,
with the decoration \texttt{(1)} corresponding to the keyword \texttt{const}.
The apparent specification $\Tbank_\app$ may be recovered from $\Tbank_\deco$ 
by dropping the decorations. 
In addition, we claim that the intended semantics 
can be seen as a \emph{decorated model} of this decorated signature:
this will become clear in Section~\ref{subsec:deco}. 
In order to add to the signature constants of type \texttt{int} 
like \texttt{0}, \texttt{1}, \texttt{2}, \dots and 
the usual operations on integers, 
a third decoration is used: 
the decoration \texttt{(0)} for \emph{pure} functions, 
which means, for functions which neither inspect nor modify 
the state of the bank account. 
So, we add to the apparent and explicit signatures the constants 
$\texttt{0},\;\texttt{1},\;$\dots$:\texttt{void}\to\texttt{int}$ 
and the operations $\texttt{+},\;\texttt{-},\;\texttt{$\ast$}: 
\texttt{int}\times\texttt{int} \to\texttt{int}$, 
and we add to the decorated signature the pure constants 
$\texttt{0}^\texttt{(0)},\;\texttt{1}^\texttt{(0)},\;$\dots$:
\texttt{void}\to\texttt{int}$
and the pure operations 
$\texttt{+}^\texttt{(0)},\;\texttt{-}^\texttt{(0)},\texttt{$\ast$}^\texttt{(0)}: 
\texttt{int}\times\texttt{int} \to\texttt{int}$.
For instance the \cpp expressions  
  $ \texttt{deposit(7); balance()} $
and 
  $\texttt{7 + balance()} $ 
can be seen as the decorated terms:
  $$ \texttt{balance}^\texttt{(1)}\circ \texttt{deposit}^\texttt{(2)}\circ 
  \texttt{7}^\texttt{(0)} 
  \quad\mbox{ and }\quad 
  \texttt{+}^\texttt{(0)}\circ 
  \tuple{\texttt{7}^\texttt{(0)},\texttt{balance}^\texttt{(1)}} $$
which may be illustrated as: 
 $$ \begin{array}{l}
  \xymatrix@C=6pc{
  \texttt{void} \ar[r]^{\texttt{7}^\texttt{(0)}} & 
  \texttt{int} \ar[r]^{\texttt{deposit}^\texttt{(2)}} & 
  \texttt{void} \ar[r]^{\texttt{balance}^\texttt{(1)}} & 
  \texttt{int} \\ } \\ 
  \mbox{ and }\quad 
  \xymatrix@C=6pc{
  \texttt{void} \ar[r]^{\tuple{\texttt{7}^\texttt{(0)},
    \texttt{balance}^\texttt{(1)}}\;} & 
  \texttt{int}\times\texttt{int} \ar[r]^{\texttt{+}^\texttt{(0)}} & 
  \texttt{int} \\ } \\ 
  \end{array} $$ 
These two decorated terms have different effects: 
the first one does modify the state while the second one is an accessor;
however, both return the same integer.
Let us introduce the symbol $\eqw$ for the relation ``same result,
maybe distinct effects''. Then: 
$$\texttt{balance}^\texttt{(1)}\circ \texttt{deposit}^\texttt{(2)}\circ 
  \texttt{7}^\texttt{(0)}
\qquad\eqw\qquad 
\texttt{+}^\texttt{(0)} \circ
  \tuple{\texttt{7}^\texttt{(0)},\texttt{balance}^\texttt{(1)}}$$

\subsection{Diagrammatic logics}
\label{subsec:dialog}

In this paper, in order to deal with a relevant notion of morphisms 
between logics, we define a \emph{logic} as a \emph{diagrammatic logic},
in the sense of \cite{DD10}. 
For the purpose of this paper let us simply say that a logic $\logL$
determines a category of theories $\catT$ which is cocomplete,  
and that a morphism of logics is a left adjoint functor,
so that it preserves the colimits. 
The objects of $\catT$ are called the a \emph{theories} of the logic $\logL$.
Quite often, $\catT$ is a category of structured categories.
The \emph{inference rules} of the logic $\logL$ describe the structure
of its theories. 
When a theory $\Ff$ is generated by some presentation  
or \emph{specification} $\Ss$, 
a \emph{model} of $\Ss$ with values in a theory $\Tt$ 
is a morphism $M:\Ff\to\Tt$ in $\catT$. 

\paragraph{The monadic equational logic.}

For instance, and for future use in the paper, 
here is the way we describe the \emph{monadic equational logic}
$\logL_{meqn}$. 
In order to focus on the syntactic aspect of the theories,
we use a congruence symbol ``$\equiv$'' rather than the equality symbol ``$=$''.
Roughly speaking, a monadic equational theory is a sort of category
where the axioms hold only up to congruence (in fact, it is a 2-category). 
Precisely, a \emph{monadic equational theory} is a directed graph 
(its vertices are called \emph{objects} or \emph{types} 
and its edges are called \emph{morphisms} or \emph{terms})
with an \emph{identity} term $\id_X:X\to X$ for each type $X$ 
and a \emph{composed} term $g\circ f:X\to Z$ for each pair 
of consecutive terms $(f:X\to Y,g:Y\to Z)$;
in addition it is endowed with \emph{equations} $f\equiv g:X\to Y$
which form a \emph{congruence}, which means, 
an equivalence relation on parallel terms compatible with the composition;
this compatibility can be split in two parts: 
\emph{substitution} and \emph{replacement}.  
In addition, the associativity and identity axioms hold up to congruence.
These properties of the monadic equational theories 
can be described by a set of \emph{inference rules},
as in Figure~\ref{fig:meqn}. 

\begin{figure}[!ht]
\renewcommand{\arraystretch}{2.2}
$$ \begin{array}{|c|} 
\hline 
\rnid \dfrac{X}{\id_X:X\to X } \qquad 
\rncomp \dfrac{f:X\to Y \quad g:Y\to Z}{g\circ f:X\to Z} \\ 
\rnidsrc \dfrac{f:X\to Y}{f\circ \id_X \equiv f} \qquad 
\rnidtgt \dfrac{f:X\to Y}{\id_Y\circ f \equiv f} \qquad
\rnassoc \dfrac{f:X\to Y \quad g:Y\to Z \quad h:Z\to W}
  {h\circ (g\circ f) \equiv (h\circ g)\circ f} \\
\rnsrefl \dfrac{}{f \equiv f} \qquad 
\rnssym \dfrac{f \equiv g}{g \equiv f} \qquad 
\rnstrans\dfrac{f \equiv g \quad g \equiv h}{f \equiv h} \\ 
\rnssubs \dfrac{f:X\to Y \quad g_1\equiv g_2:Y\to Z}
  {g_1\circ f \equiv g_2\circ f :X\to Z}  \qquad 
\rnsrepl \dfrac{f_1\equiv f_2:X\to Y \quad g:Y\to Z}
  {g\circ f_1 \equiv g\circ f_2 :X\to Z} \\
\hline 
\end{array}$$
\renewcommand{\arraystretch}{1}
\caption{Rules of the monadic equational logic}
\label{fig:meqn}
\end{figure}

\paragraph{Adding products to the monadic equational logic.}

In contrast with equational theories, 
the existence of products is not required in a monadic equational theory.
However some specific products may exist. 
A product in a monadic equational theory $\catT$ 
is ``up to congruence'', in the following sense. 
Let $(Y_i)_{i\in I}$ be a family of objects in $\catT$,
indexed by some set $I$. 
A \emph{product} with base $(Y_i)_{i\in I}$ is a cone 
$(q_i:Y\to Y_i)_{i\in I}$ such that for every cone 
$(f_i:X\to Y_i)_{i\in I}$ on the same base there is a term 
$f=\tuple{f_i}_{i\in I}:X\to Y$ such that $q_i\circ f\equiv f_i$ for each $i$,
and in addition this term is unique up to congruence,
in the sense that 
if $g:X\to Y$ is such that $q_i\circ g\equiv f_i$ for each $i$ 
then $g\equiv f$. 
When $I$ is empty, we get a \emph{terminal} object $\unit$,
such that for every $X$ there is an arrow $\tu_X:X\to \unit$  
which is unique up to congruence. 
The corresponding inference rules are given in Figure~\ref{fig:meqn-prod}. 
The quantification ``$\forall i$'',
or ``$\forall i\in I$'', is a kind of ``syntactic sugar'':
when occuring in the premisses of a rule,  
it stands for a conjunction of premisses. 

\begin{figure}[!ht] 
\renewcommand{\arraystretch}{1.8}
$$ \begin{array}{|c|} 
\hline 
\multicolumn{1}{|l|}
  {\text{When $(q_i:Y\to Y_i)_{i\in I}$ is a product:}} \\ 
\rntuple \dfrac {(f_i\!:\!X\to Y_i)_i} 
  {\tuple{f_i}_i\!:\!X\to Y} \quad 
\rnproj \dfrac {(f_i\!:\!X\to Y_i)_i} 
  {q_i \circ \tuple{f_j}_j \equiv f_i } \quad 
\rntupleun \dfrac{g:X\to Y \quad \forall i\; q_i \circ g \equiv f_i}
  {g \equiv \tuple{f_j}_j} \\
\multicolumn{1}{|l|}
  {\text{When $\unit$ is a terminal type (``empty product''):}} \\ 
\rnfinal \dfrac{X}{\tu_X:X\to \unit} \qquad  
\rnfinalun \dfrac{g:X\to \unit}
  {g \equiv \tu_X}  \\\hline 
\end{array}$$
\renewcommand{\arraystretch}{1}
\caption{Rules for products} 
\label{fig:meqn-prod} 
\end{figure}

\section{Three logics for states}  
\label{sec:states}

In this section we introduce three logics for dealing with states 
as computational effects.
This generalizes 
the example of the bank account in Section~\ref{subsec:bank}.
We present first the explicit logic (close to the semantics),
then the apparent logic (close to the syntax), 
and finally the decorated logic and the morphisms from 
the decorated logic to the apparent and the explicit ones.
In the syntax of an imperative language  
there is no type of states (the state is ``hidden'')  
while the interpretation of this language 
involves a set of states $\St$. 
More precisely, if the types $X$ and $Y$ 
are interpreted as the sets $\deno{X}$ and $\deno{Y}$, 
then each term $f:X\to Y$ is interpreted 
as a function $\deno{f}:\deno{X}\times \St \to \deno{Y}\times \St$. 
In Moggi's paper introducing monads for effects \cite{Mo91}
such a term $f:X\to Y$ is called a \emph{computation},
and whenever the function $\deno{f}$ is $\deno{f}_0\times\id_{\St}$ 
for some $\deno{f}_0:\deno{X} \to \deno{Y}$ then $f$ is called 
a \emph{value}.
We keep this distinction, using \emph{modifier} and \emph{pure term}
instead of \emph{computation} and \emph{value}, respectively.
In addition, an \emph{accessor}
(or \emph{inspector}) is a term $f:X\to Y$ that is interpreted by a function 
$\deno{f}=\tuple{\deno{f}_1,q_X}$,   
for some $\deno{f}_1:\deno{X}\times \St \to \deno{Y}$, 
where $q_X:\deno{X}\times \St \to \St$ is the projection.   
It follows that every pure term is an accessor and every accessor is 
a modifier. 
We will respectively use the decorations $\pure$, $\acc$ and $\modi$, 
written as superscripts, 
for pure terms, accessors and modifiers.
Moreover, we distinguish two kinds of equations:
when $f,g:X\to Y$ are parallel terms, then 
a \emph{strong} equation $f \eqs g$ is interpreted as the equality 
$\deno{f}=\deno{g}: \deno{X}\times \St \to \deno{Y}\times \St$,
while a \emph{weak} equation $f \eqw g$ 
is interpreted as the equality 
$p_Y\circ\deno{f}=p_Y\circ\deno{g}: 
\deno{X}\times \St \to \deno{Y}$,
where $p_Y:\deno{Y}\times \St \to \deno{Y}$ is the projection.  
Clearly, strong and weak equations coincide on accessors 
and on pure terms, while they differ on modifiers. 
As in Section~\ref{subsec:states},  
we consider some given set of locations $\Loc$ 
and for each location $i$ a set $\Val_i$ of possible values for $i$.
The \emph{set of states} is defined as $\St=\prod_{i\in\Loc}\Val_i$, 
and the projections are denoted by $\sfl_{i,1}:\St\to\Val_i$. 
For each location $i$, let $\sfu_i:\Val_i\times\St\to\St$ 
be defined by Equations (1) as in Section~\ref{subsec:states}. 
In order to focus on the fundamental properties of states as effects, 
the three logics for states are based on the ``poor''
monadic equational logic (as described in Section~\ref{subsec:dialog}). 

\subsection{The explicit logic for states}
\label{subsec:expl}

The \emph{explicit logic for states} $\logL_\expl$ 
is a kind of ``pointed'' monadic equational logic: 
a theory $\Tt_\expl$ for $\logL_\expl$ 
is a monadic equational theory with  
a distinguished object $S$, called the \emph{type of states},
and with a product-with-$S$ functor $X\times S$.
As in Section~\ref{subsec:bank}, the explicit logic provides 
the relevant semantics, but it is far from the syntax.
The explicit theory for states $\Tstate_\expl$ is generated by 
a type $V_i$ and    
an operation $l_{i,1} : S\to V_i$ for each location $i$, 
which form a product $(l_{i,1} :S\to V_i)_{i\in\Loc}$.
Thus, for each location $i$ there is an operation 
$u_i:V_i\times S\to S$, unique up to congruence, 
which satisfies the equations below
(where $p_i: V_i\times S\to V_i$ and $q_i: V_i\times S\to S$ 
are the projections): 
  $$ \Tstate_\expl : 
	\begin{cases}
	\mbox{ operations } & 
	  l_{i,1} : S\to V_i \;,\; 
	  u_i:V_i\times S\to S \\ 
  \mbox{ product } & 
	  (l_{i,1} :S\to V_i)_{i\in\Loc} \\ 
	\mbox{ equations } & 
	  l_{i,1}\circ u_i \eqs p_i : V_i\times S\to V_i \;,\; 
    l_{j,1}\circ u_i \eqs l_{j,1} \circ q_i : V_i\times S\to V_j \; 
		\mbox{ for each }j\ne i \\
   \end{cases}$$
Let us define the explicit theory $\Tset_\expl$ 
as the category of sets with the equality as congruence 
and with the set of states $\St=\prod_{j\in\Loc} \Val_j$
as its distinguished set. 
The semantics of states, as described in Section~\ref{subsec:states},
is the model $M_\expl:\Tstate_\expl\to\Tset_\expl$ 
which maps the type $V_i$ to the set $\Val_i$ for each $i\in\Loc$, 
the type $S$ to the set $\St$, 
and the operations $l_{i,1}$ and $u_i$
to the functions $\sfl_{i,1}$ and $\sfu_i$, respectively. 

\subsection{The apparent logic for states}
\label{subsec:app}

The \emph{apparent logic for states} $\logL_{\app}$ 
is the monadic equational logic (Section~\ref{subsec:dialog}).
As in Section~\ref{subsec:bank}, the apparent logic is close 
to the syntax but it does not provide the relevant semantics. 
The \emph{apparent theory for states} $\Tstate_{\app}$ 
can be obtained from the explicit theory $\Tstate_{\expl}$ 
by identifying the type of states $S$ with the unit type $\unit$. 
So, there is in $\Tstate_{\app}$ a terminal type $\unit$ and 
for each location $i$ a type $V_i$ for the possible values of $i$ 
and an operation $l_i:\unit \to V_i$ for observing the value of $i$.
A set-valued model for this part of $\Tstate_{\app}$, 
with the constraint that for each $i$ the interpretation of $V_i$ 
is the given set $\Val_i$,
is made of an element $a_i\in\Val_i$  for each $i$
(it is the image of the interpretation of $l_i$). 
Thus, such a model 
corresponds to a state, made of a value for each location; 
this is known as the \emph{states-as-models} or \emph{states-as-algebras}  
point of view \cite{GDK96}. 
In addition, it is assumed that in $\Tstate_{\app}$ 
the operations $l_i$'s form a product $(l_i:\unit\to V_i)_{i\in\Loc}$. 
This assumption implies that each $l_i$ is an isomorphism,
so that each $V_i$ must be interpreted as a singleton:
this does not fit with the semantics of states. 
However, we will see in Section~\ref{subsec:deco} 
that this assumption becomes meaningful when decorations are added,
in a similar way as in the bank example in Section~\ref{subsec:bank}. 
Formally, the assumption that $(l_i:\unit\to V_i)_{i\in\Loc}$
is a product provides for each location $i$ an operation 
$u_i:V_i\to \unit$, unique up to congruence, 
which satisfies the equations below
(where $\id_i:V_i\to V_i$ is the identity and $\tu_i=\tu_{V_i}:V_i\to\unit$)~:
	 $$ \Tstate_\app : 
	\begin{cases}
	\mbox{ operations } & 
	  l_i : \unit\to V_i \;,\; 
	  u_i:V_i \to \unit \\ 
  \mbox{ product } & 
	  (l_i : \unit\to V_i)_{i\in\Loc} 
		\;\; \mbox{ with terminal type } \unit \\ 
	\mbox{ equations } & 
	  l_i\circ u_i \eqs \id_i : V_i\to V_i \;,\; 
    l_j\circ u_i \eqs l_j\circ \tu_i : V_i\to V_j  \; 
		\mbox{ for each }j\ne i \\
   \end{cases}$$
At first view, these equations mean that after $u_i(a)$ is executed, 
the value of $i$ is put to $a$ and the value of $j$ (for $j\ne i$) 
is unchanged. However, as noted above, 
this intuition is not supported by the semantics in the apparent logic.
However, the apparent logic can be used for checking the validity 
of a decorated proof, as explained in Section~\ref{subsec:deco-app}.

\subsection{The decorated logic for states}
\label{subsec:deco}

Now, as in Section~\ref{subsec:bank}, 
we introduce a third logic for states, which is 
close to the syntax and which provides the relevant semantics. 
It is defined by adding ``decorations'' to the apparent logic. 
A theory $\Tt_{\deco}$ for the \emph{decorated logic for states} 
$\logL_{\deco}$ is made of:
\begin{itemize}
\item A monadic equational theory $\Tt^\modi$.
The terms in $\Tt^\modi$ may be called the \emph{modifiers}  
and the equations $f \eqs g$ may be called the \emph{strong equations}. 
\item Two additional monadic equational theories 
$\Tt^\pure$ and $\Tt^\acc$, with the same types as $\Tt^\modi$, 
and such that $\Tt^\pure \subseteq \Tt^\acc \subseteq \Tt^\modi$ 
and the congruence on $\Tt^\pure$ and on $\Tt^\acc$
is the restriction of the congruence on $\Tt^\modi$. 
The terms in $\Tt^\acc$ may be called the \emph{accessors}, 
and if they are in $\Tt^\pure$ they may be called the \emph{pure terms}. 
\item A second equivalence relation $\eqw$ between parallel terms 
in $\Tt^\modi$, 
which is only ``weakly'' compatible with the composition; 
the relation $\eqw$ satisfies the substitution property 
but only a weak version of the replacement property,
called the \emph{pure replacement}: 
if $f_1\eqw f_2:X\to Y$ and $g:Y\to Z$ 
then in general $g\circ f_1 \not\eqw g\circ f_2$,
except when $g$ is pure. 
The relations $f \eqw g$ are called the \emph{weak equations}.
It is assumed that every strong equation is a weak equation and  
that every weak equation between accessors is a strong equation,
so that the relations $\eqs$ and $\eqw$ coincide on $\Tt^\pure$ 
and on $\Tt^\acc$.
\end{itemize}
We use the following notations, called \emph{decorations}: 
a pure term $f$ is denoted $f^\pure$, 
an accessor $f$ is denoted $f^\acc$, 
and a modifier $f$ is denoted $f^\modi$; 
this last decoration is unnecessary since every term is a modifier, 
however it may be used for emphasizing.
Figure~\ref{fig:deco} provides the \emph{decorated rules}, 
which describe the properties of the decorated theories. 
For readability, the decoration properties may be grouped 
with other properties: for instance, ``$f^\acc \eqw g^\acc$''
means ``$f^\acc$ and $g^\acc$ and $f \eqw g$''. 

\begin{figure}[!ht]
\renewcommand{\arraystretch}{2.8}
$$ \begin{array}{|c|} 
\hline 
\multicolumn{1}{|l|}{ \txt{Rules of the monadic equational logic, and: }} \\ 
\rnpid \dfrac{X}{\id_X^\pure:X\to X } \qquad 
\rnpcomp \dfrac{f^\pure \quad g^\pure}{(g\circ f)^\pure}  \qquad 
\rnpa \dfrac{f^\pure}{f^\acc} \qquad
\rnacomp \dfrac{f^\acc \quad g^\acc}{(g\circ f)^\acc} \\ 
\rnws \dfrac{f^\acc \eqw g^\acc}{f \eqs g} \qquad
\rnsw \dfrac{f \eqs g}{f \eqw g} \\ 
\rnwrefl \dfrac{}{f \eqw f} \qquad 
\rnwsym \dfrac{f \eqw g}{g \eqw f} \qquad 
\rnwtrans \dfrac{f \eqw g \quad g \eqw h}{f \eqw h} \\ 
\rnwsubs \dfrac{f:X\to Y \quad g_1\eqw g_2:Y\to Z}
  {g_1\circ f \eqw g_2\circ f :X\to Z}  \qquad 
\rnwrepl \dfrac{f_1\eqw f_2:X\to Y \quad g^\pure:Y\to Z}
  {g\circ f_1 \eqw g\circ f_2 :X\to Z} \\
\hline 
\end{array}$$
\renewcommand{\arraystretch}{1}
\caption{Rules of the decorated logic for states}
\label{fig:deco}
\end{figure}

Some specific kinds of products may be used in a decorated theory,
for instance:
\begin{itemize}
\item A distinguished type $\unit$  
with the following \emph{decorated terminality} property: 
for each type $X$ there is a pure term $\tu_X:X\to \unit$ 
such that every modifier $g:X\to \unit$ satisfies $g \eqw \tu_X$. 
It follows from the properties of weak equations 
that $\unit$ is a terminal type in $\Tt^\pure$ and in $\Tt^\acc$.
\item An \emph{observational product} 
with base $(Y_i)_{i\in I}$ is a cone of accessors 
$(q_i:Y\to Y_i)_{i\in I}$ such that for every cone of accessors 
$(f_i:X\to Y_i)_{i\in I}$ on the same base there is a modifier 
$f=\tuple{f_i}_{i\in I}:X\to Y$ such that 
$q_i\circ f\eqw f_i$ for each $i$,
and in addition this modifier is unique up to strong equations, 
in the sense that 
if $g:X\to Y$ is a modifier such that $q_i\circ g\eqw f_i$ for each $i$ 
then $g\eqs f$. 
An observational product allows to prove strong equations from weak ones:
by looking at the results of some observations, 
thanks to the properties of the observational product, 
we get information on the state. 
\end{itemize}

\begin{figure}[!ht] 
\renewcommand{\arraystretch}{2.3}
$$ \begin{array}{|c|} 
\hline 
\multicolumn{1}{|l|}
  {\text{When $\unit$ is a decorated terminal type:}} \\ 
\rnpfinal \dfrac{X}{\tu_X^\pure:X\to \unit} \qquad  
\rnwfinalun \dfrac{g:X\to \unit}{g \eqw \tu_X} \\ 
\multicolumn{1}{|l|}
  {\text{When $(q_i^\acc:Y\to Y_i)_i$ is an observational product:}
  \quad
\rnatuple \dfrac {(f_i^\acc\!:\!X\to Y_i)_i} 
  {\tuple{f_i}_i^\modi\!:\!X\to Y} \;\;  
}\\ 
\rnaproj \dfrac {(f_i^\acc\!:\!X\to Y_i)_i} 
  {q_i \circ \tuple{f_j}_j \eqw f_i } \qquad
\rnstupleun  \dfrac{g^\modi\!:\!X\to Y \;\; 
  \forall i \; q_i \circ g \eqw f_i^\acc}
  {g \eqs \tuple{f_j}_j}  \\
\hline 
\end{array}$$
\renewcommand{\arraystretch}{1}
\caption{Rules for some decorated products for states} 
\label{fig:deco-prod} 
\end{figure}

The decorated theory of states $\Tstate_\deco$ is generated by 
a type $V_i$  
and an accessor $l_i^\acc:\unit\to V_i$ for each $i\in\Loc$, 
which form an observational product 
$(l_i^\acc:\unit\to V_i)_{i\in\Loc}$. 
The modifiers $u_i$'s are defined (up to strong equations), 
using the property of the observational product, 
by the weak equations below:
	$$ \Tstate_\deco : 
	\begin{cases}
	\mbox{ operations } & 
	  l_i^\acc : \unit\to V_i \;,\; 
	  u_i^\modi : V_i \to \unit \\ 
  \mbox{ observational product } & 
	  (l_i^\acc : \unit\to V_i)_{i\in\Loc} 
		\;\; \mbox{ with decorated terminal type } \unit \\ 
	\mbox{ equations } & 
	  l_i \circ u_i \eqw \id_i : V_i\to V_i  \;,\; 
		l_j \circ u_i \eqw l_j \circ \tu_i : V_i\to V_j \; 
		\mbox{ for each }j\ne i \\
   \end{cases}$$

The decorated theory of sets $\Tset_\deco$ is built from the category of sets, 
as follows. 
There is in $\Tset_\deco$ 
a type for each set, 
a modifier $f^\modi:X\to Y$ for each function $f:X\times \St \to Y\times \St$,
an accessor $f^\acc:X\to Y$ for each function $f:X\times \St \to Y$,  
and a pure term $f^\pure:X\to Y$ for each function $f:X \to Y$,
with the straightforward conversions.  
Let $f^\modi,g^\modi:X\to Y$ corresponding to $f,g:X\times \St \to Y\times \St$. 
A strong equation $f\eqs g$ is an equality $f=g:X\times\St \to Y\times\St$,
while a weak equation $f\eqw g$ 
is an equality $p\circ f=p\circ g:X\times\St \to Y$,
where $p:Y\times\St \to Y$ is the projection.  
For each location $i$ the projection $\sfl_i:\St\to\Val_i$ 
corresponds to an accessor $\sfl_i^\acc:\unit\to\Val_i$ in $\Tset_\deco$,
so that the family $(\sfl_i^\acc)_{i\in\Loc}$ forms  
an observational product in $\Tset_\deco$. 
We get a model $M_\deco$ of $\Tstate_\deco$ with values in $\Tset_\deco$ 
by mapping 
the type $V_i$ to the set $\Val_i$ and 
the accessor $l_i^\acc$ to the accessor $\sfl_i^\acc$, for each $i\in\Loc$.
Then for each $i$ 
the modifier $u_i^\modi$ is mapped to the modifier $\sfu_i^\modi$. 

\subsection{From decorated to apparent}
\label{subsec:deco-app}

Every decorated theory $\Tt_\deco$ gives rise to an apparent theory 
$\Tt_\app$ by dropping the decorations, which means that 
the apparent theory $\Tt_\app$ is made of 
 a type $X$ for each type $X$ in $\Tt_\deco$, 
  a term $f:X\to Y$ for each modifier $f:X\to Y$ in $\Tt_\deco$
  (which includes the accessors and the pure terms), 
  and an equation $f\equiv g$ for each weak equation $f\eqw g$ in $\Tt_\deco$
  (which includes the strong equations). 
Thus, the distinction between modifiers, accessors and pure terms disappears,
as well as the distinction between weak and strong equations. 
Equivalently, the apparent theory $\Tt_\app$ can be defined 
as the apparent theory $\Tt^\modi$ together with an 
equation $f\equiv g$ for each weak equation $f\eqw g$ in $\Tt_\deco$
which is not associated to a strong equation in $\Tt_\deco$
(otherwise, it is yet in $\Tt^\modi$). 
Thus, a decorated terminal type in $\Tt_\deco$ 
becomes a terminal type in $\Tt_\app$ 
and an observational product $(q_i^\acc:Y\to Y_i)_i$ in $\Tt_\deco$ 
becomes a product $(q_i:Y\to Y_i)_i$ in $\Tt_\app$.
In the same way, each rule of the decorated logic is mapped 
to a rule of the apparent logic by dropping the decorations. 
This property can be used for checking a decorated proof in two steps, 
by checking on one side the undecorated proof 
and on the other side the decorations. 
This construction of $\Tt_\app$ from $\Tt_\deco$,
by dropping the decorations, is a morphism from 
$\logL_\deco$ to $\logL_\app$, denoted $F_\app$.

\subsection{From decorated to explicit}
\label{subsec:deco-expl}

Every decorated theory $\Tt_\deco$ gives rise to an explicit theory 
$\Tt_\expl$ by \emph{expanding} the decorations, which means 
that the explicit theory $\Tt_\expl$ is made of:
  \begin{itemize}
  \item A type $X$ for each type $X$ in $\Tt_\deco$;
  projections are denoted by
  $p_X:X\times S \to X$ and $q_X:X\times S \to S$.
  \item A term $f:X\times S\to Y\times S$ 
  for each modifier $f:X\to Y$ in $\Tt_\deco$, such that: 
    \begin{itemize}
    \item if $f$ is an accessor 
    then there is a term $f_1:X\times S\to Y$ in $\Tt_\expl$
    such that $f=\tuple{f_1,q_X}$,
    \item if moreover $f$ is a pure term  
    then there is a term $f_0:X\to Y$ in $\Tt_\expl$ 
    such that $f_1=f_0\circ p_X: X\times S\to Y$,
    hence $f=\tuple{f_0\circ p_X,q_X}=f_0\times\id_S$ in $\Tt_\expl$.
    \end{itemize}
  \item An equation $f \equiv g : X\times S\to Y\times S$ 
  for each strong equation $f \eqs g : X\to Y$ in $\Tt_\deco$. 
  \item An equation $p_Y \circ f \equiv p_Y \circ g : 
  X\times S\to Y$ 
  for each weak equation $f \eqw g : X\to Y$ in $\Tt_\deco$. 
  \item A product $(q_{i,1}:Y\times S\to Y_i)_i$ 
  for each observational product $(q_i^\acc:Y\to Y_i)_i$ 
  in $\Tt_\deco$.  
  \end{itemize}
This construction of $\Tt_\expl$ from $\Tt_\deco$ is a morphism from 
$\logL_\deco$ to $\logL_\expl$, denoted $F_\expl$
and called the \emph{expansion}. 
The expansion morphism makes explicit the meaning of the decorations,
by introducing a ``type of states'' $S$.
Thus, each modifier $f^\modi$ gives rise to a term $f$ 
which may use and modify the state,
while whenever $f^\acc$ is an accessor then $f$ may use the state 
but is not allowed to modify it,
and when moreover $f^\pure$ is a pure term then $f$ 
may neither use nor modify the state.
When $f^\modi \eqs g^\modi$ then $f$ and $g$ 
must return the same result and the same state;  
when $f^\modi \eqw g^\modi$ then $f$ and $g$ 
must return the same result but maybe not the same state. 
We have seen that the semantics of states
cannot be described in the apparent logic, 
but can be described both in the 
decorated logic and in the explicit logic. 
It should be reminded that every morphism of logics is a left adjoint functor.  
This is the case for the expansion morphism $F_\expl:\logL_\deco\to\logL_\expl$:
it is a left adjoint functor $F_\expl:\catT_\deco\to\catT_\expl$, 
its right adjoint is denoted $G_\expl$. 
In fact, it is easy to check that $\Tset_\deco = G_\expl(\Tset_\expl)$,
and since $\Tstate_\expl = F_\expl(\Tstate_\deco)$ it follows that  
the decorated model $M_\deco:\Tstate_\deco\to\Tset_\deco$ 
and the explicit model $M_\expl:\Tstate_\expl\to\Tset_\expl$
are related by the adjunction $F_\expl \dashv G_\expl$. 
This means that the models $M_\deco$ and $M_\expl$ are two different 
ways to formalize the semantics of states from Section~\ref{subsec:states}.
In order to conclude Section~\ref{sec:states}, 
the morphims of logic $F_\app$ and $F_\expl$ are summarized 
in Figure~\ref{fig:span}. 

\begin{figure}[!ht] 
\renewcommand{\arraystretch}{1.1}
$$ \begin{array}{|l|ll|l|}
\hline 
\multicolumn{4}{|c|}{
\xymatrix@C=9pc{\Tt_\app & 
\ar[l]_{F_\app} \Tt_\deco \ar[r]^{F_\expl} & 
\Tt_\expl \\ }
} \\ 
\hline 
f:X\to Y & 
\mbox{ modifier } &
f:X\to Y  & 
f:X\times S\to Y\times S  \\
f:X\to Y & 
\mbox{ accessor } &
f^\acc:X\to Y  & 
f_1:X\times S\to Y  \\
f:X\to Y & 
\mbox{ pure term } &
f^\pure:X\to Y  &
f_0:X\to Y  \\
\hline 
f \eqs g:X\to Y &
\mbox{ strong equation } &
f \eqs g:X\to Y &
f \eqs g : X\times S\to Y\times S  \\
f \eqs g:X\to Y &
\mbox{ weak equation } &
f \eqw g:X\to Y &
p_Y\circ f \eqs p_Y\circ g : X\times S\to Y \\
\hline 
\end{array} $$
\renewcommand{\arraystretch}{1}
\caption{A span of logics for states} 
\label{fig:span} 
\end{figure}

\section{Decorated proofs} 
\label{sec:proofs}

The inference rules of the decorated logic $\logL_\deco$
are now used for proving some of the Equations (2)  
(in Remark~\ref{rem:states-eq}).
All proofs in this section are performed in the decorated logic; 
for readability the identity and associativity rules  
$\rnidsrc$, $\rnidtgt$ and $\rnassoc$ are omitted. 
Some derived rules are proved in Section~\ref{subsec:lemmas}, 
then Equation (2.1) is proved in Section~\ref{subsec:liui}.
In order to deal with the equations with two values as argument or as result,
we use the semi-pure products introduced in \cite{DDR11}; 
the rules for semi-pure products are reminded 
in Section~\ref{subsec:prod}, then all seven Equations (2)
are expressed in the decorated logic
and Equation (2.6) is proved in Section~\ref{subsec:uiuj}.
Proving the other equations would be similar.
We use as axioms
the fact that $l_i$ is an accessor and the weak equations 
in $\Tstate_\deco$ (Section~\ref{subsec:deco}). 

\subsection{Some derived rules} 
\label{subsec:lemmas} 

Let us now derive some rules from the rules of the decorated logic 
(Figures~\ref{fig:deco} and~\ref{fig:deco-prod}). 

\begin{figure}[!ht]
\renewcommand{\arraystretch}{2.5}
\small
$$ \begin{array}{|c|c|} 
\hline 
(E_1^\acc) \;\;
  \dfrac{f^\acc:X\to \unit }{f \eqs \tu_X} &  
(E_1^\pure) \;\;
  \dfrac{f^\pure:X\to \unit }{f \eqs \tu_X} \\ 
(E_2^\acc) \;\;
  \dfrac{f^\acc: X \to \unit \quad g^\acc: X \to \unit}{f \eqs g } & 
(E_2^\pure) \;\;
  \dfrac{f^\pure: X \to \unit \quad g^\pure: X \to \unit}{f \eqs g } \\ 
(E_3^\acc) \;\;
  \dfrac{f^\acc: X \to Y \quad g^\acc: Y \to \unit \quad 
    h^\acc: X \to \unit}{g\circ f \eqs h} & 
(E_3^\pure) \;\;
  \dfrac{f^\pure: X \to Y \quad g^\pure: Y \to \unit \quad 
    h^\pure: X \to \unit}{g\circ f \eqs h} \\ 
(E_4^\acc) \;\;
  \dfrac{f^\acc: \unit \to X}{\tu_{X} \circ f \eqs \id_\unit } & 
(E_4^\pure) \;\;
  \dfrac{f^\pure: \unit \to X}{\tu_{X} \circ f \eqs \id_\unit } \\ 
\hline 
\end{array}$$
\normalsize
\renewcommand{\arraystretch}{1}
\caption{Some derived rules in the decorated logic for states}
\label{fig:derived}
\end{figure}

\proof
The derived rules in the left part of Figure~\ref{fig:derived} 
can be proved as follows.   
The proof of the rules in the right part are left to the reader.
\small
\begin{prooftree}
\AxiomC{$f^\acc$}
      \AxiomC{$X$}
      \LeftLabel{\rnpfinal} 
      \UnaryInfC{$\tu_X^\pure$}
      \LeftLabel{\rnpa} 
      \UnaryInfC{$\tu_X^\acc$}
          \AxiomC{$f:X\to \unit$}
          \LeftLabel{\rnwfinalun} 
          \UnaryInfC{$f \eqw \tu_X$}
      \LeftLabel{\rnws} 
      \TrinaryInfC{$f \eqs \tu_X \;\; (E_1^\acc) $}
\end{prooftree}
$$
\AxiomC{$f^\acc\!:\!\unit \!\to\! X$}
\LeftLabel{$(E_1^\acc)$} 
\UnaryInfC{$f \eqs \tu_\unit$}
      \AxiomC{$g^\acc\!:\!\unit \!\to\! X$}
      \LeftLabel{$(E_1^\acc)$} 
      \UnaryInfC{$g \eqs \tu_\unit$}
      \LeftLabel{\rnsym} 
      \UnaryInfC{$ \tu_\unit \eqs  g$}
   \LeftLabel{\rntrans} 
   \BinaryInfC{$ f \eqs g \;\; (E_2^\acc)$}
\DisplayProof 
\qquad
\AxiomC{$f^\acc\!:\!X \!\to\! Y$}
    \AxiomC{$g^\acc\!:\!Y \!\to\! \unit$}
  \LeftLabel{$\rnacomp$} 
  \BinaryInfC{$(g\!\circ\! f)^\acc\!:\!X \!\to\! \unit$}
      \AxiomC{$h^\acc\!:\!X \!\to\! \unit$}
      \LeftLabel{$(E_2^\acc)$} 
      \BinaryInfC{$g\!\circ\! f \eqs h \;\; (E_3^\acc)$}
\DisplayProof $$
\begin{prooftree}
\AxiomC{$f^\acc:\unit \to X$}
    \AxiomC{$X$}
    \LeftLabel{\rnpfinal} 
    \UnaryInfC{$\tu_X^\pure:X\to\unit$}
    \LeftLabel{\rnpa} 
    \UnaryInfC{$\tu_X^\acc:X\to\unit$}
        \AxiomC{$\unit$}
        \LeftLabel{\rnpid} 
        \UnaryInfC{$\id_\unit^\pure:\unit\to\unit$}
        \LeftLabel{\rnpa} 
        \UnaryInfC{$\id_\unit^\acc:\unit\to\unit$}
    \LeftLabel{$(E_3^\acc)$} 
    \TrinaryInfC{$\tu_X \circ f \eqs \id_\unit \;\; (E_4^\acc)$}
\end{prooftree}
\normalsize 
\qed

\subsection{Annihilation lookup-update} 
\label{subsec:liui} 

It is easy to check that the 
decorated equation $u_i^\modi \circ l_i^\acc \eqs \id_\unit^\pure $ 
gets expanded as $u_i \circ l_i \equiv \id_S $,
which clearly gets interpreted as Equation (2.1) in 
Remark~\ref{rem:states-eq}. 
Let us prove this decorated equation,
using the axioms (for each location $i$),
from $\Tstate_\deco$ in Section~\ref{subsec:deco}:
  $$ 
  (A_0)\;\; l_i^\acc \;,\qquad 
  (A_1)\;\; l_i \circ u_i  \eqw \id_i \;,\qquad 
  (A_2)\;\; l_j \circ u_i  \eqw l_j \circ \tu_i \; \mbox{ for each } j\ne i \;.
  $$

\begin{prop}
\label{prop:liui}
For each location $i\,$,
reading the value of
a location $i$ and then updating the location $i$ with the
obtained value is just like doing nothing.
  $$ u_i^\modi \circ l_i^\acc \eqs \id_\unit^\pure : \unit \to \unit \;.$$
\end{prop}

\proof
Let $i$ be a location. 
Using the unicity property of the observational product, 
we have to prove that 
  $  l_k \circ u_i \circ l_i \eqw l_k : \unit \to V_k $ for each location $k\,$.
\begin{itemize}
\item 
When $k=i$, the substitution rule for $\eqw$ yields: 
\small
\begin{prooftree}  
\AxiomC{$ (A_1)\;\; l_i \circ u_i  \eqw \id_i $} 
\LeftLabel{\rnwsubs}
\UnaryInfC{$l_i \circ u_i \circ l_i \eqw l_i$}
\end{prooftree}
\normalsize
\item 
When $k\ne i$, using the substitution rule for $\eqw$ and    
the replacement rule for $\eqs$ we get: 
\small
\begin{prooftree}  
\AxiomC{$ (A_2)\;\;l_k \circ u_i  \eqw l_k \circ \tu_i $} 
\LeftLabel{\rnwsubs}
\UnaryInfC{$l_k \circ u_i \circ l_i \eqw l_k \circ \tu_i \circ l_i $}
        \AxiomC{$ (A_0)\;\; l_i^\acc$}
	\LeftLabel{$(E_4^\acc)$}
        \UnaryInfC{$\tu_i \circ l_i \eqs \id_\unit $}  
        \LeftLabel{\rnsrepl} 
        \UnaryInfC{$l_k \circ \tu_i \circ l_i \eqs l_k $} 
        \LeftLabel{\rnsw}
        \UnaryInfC{$l_k \circ \tu_i \circ l_i \eqw l_k $}
    \LeftLabel{\rnwtrans}
    \BinaryInfC{$l_k \circ u_i \circ l_i \eqw l_k$}
\end{prooftree}
\normalsize
\end{itemize}
\qed

\begin{rem}
At the top of the right branch in the proof above, 
the decoration $\acc$ for $l_i$ could not be replaced by $\modi$. 
Indeed, from $l_i^\modi$ we can derive the weak equation
$\tu_i \circ l_i \eqw \id_\unit $,
but this is not sufficient for deriving 
$l_k \circ \tu_i \circ l_i \eqw l_k $ by replacement since $l_k$ is not pure. 
\end{rem}

\subsection{Semi-pure products} 
\label{subsec:prod} 

Let $\Tt_\deco$ be a theory with respect to the decorated logic for states
and let $\Tt^\pure$ be its pure part, so that $\Tt^\pure$ 
is a monadic equational theory. 
The \emph{product} of two types $X_1$ and $X_2$ in $\Tt_\deco$
is defined as their product in $\Tt^\pure$ 
(it is a product up to strong equations, as in Section~\ref{subsec:states}). 
The projections from $X_1\times X_2$ to $X_1$ and $X_2$ are 
respectively denoted by $\pi_1^\pure$ and $\pi_2^\pure$ 
when the types $X_1$ and $X_2$ are clear from the context.
The \emph{product} of two pure morphisms $f_1^\pure:X_1\to Y_1$ 
and $f_2^\pure:X_2\to Y_2$ is a pure morphism 
$(f_1\times f_2)^\pure:X_1\times X_2\to Y_1\times Y_2$ 
subject to the rules in Figure~\ref{fig:pure-prod},
which are the usual rules for products up to strong equations.
Moreover when $X_1$ or $X_2$ is $\unit$ 
it can be proved in the usual way that the projections 
$\pi_1^\pure:X_1\times \unit \to X_1$ and 
$\pi_2^\pure:\unit\times X_2 \to X_2$ 
are isomorphisms.
The permutation $\perm_{X_1,X_2}^\pure:X_1\times X_2 \to X_2\times X_1$ 
is defined as usual by $\pi_1\circ \perm_{X_1,X_2} \eqs \pi_2$ 
and $\pi_2\circ \perm_{X_1,X_2} \eqs \pi_1$.

\begin{figure}[!ht] 
\renewcommand{\arraystretch}{2.6}
$$ \begin{array}{|c|} 
\hline 
\rnpprod \dfrac {f_1^\pure:X_1\to Y_1\quad f_2^\pure:X_2\to Y_2} 
  {(f_1\times f_2)^\pure:X_1\times X_2\to Y_1\times Y_2} \\  
\rnpprojone \dfrac {f_1^\pure:X_1\to Y_1\quad f_2^\pure:X_2\to Y_2}
  {\pi_1 \circ (f_1 \times f_2) \eqs  f_1 \circ \pi_1} \qquad 
\rnpprojtwo \dfrac {f_1^\pure:X_1\to Y_1\quad f_2^\pure:X_2\to Y_2}
  {\pi_2 \circ (f_1 \times f_2) \eqs  f_2 \circ \pi_2 } \\   
\rnpprodun \dfrac{g^\pure:X_1\times X_2\to Y_1\times Y_2 \quad 
  \pi_1 \circ g \eqs  f_1 \circ \pi_1 \quad 
  \pi_2 \circ g \eqs  f_2 \circ \pi_2 }
  {g \eqs f_1 \times f_2} \\
\hline 
\end{array}$$
\renewcommand{\arraystretch}{1}
\caption{Rules for products of pure morphisms} 
\label{fig:pure-prod} 
\end{figure}

The rules in Figure~\ref{fig:pure-prod}, 
which are symmetric in $f_1$ and $f_2$, cannot be applied to modifiers: 
indeed, the effect of building a pair of modifiers depends 
on the evaluation strategy. 
However, following \cite{DDR11}, we define the \emph{left semi-pure product} 
of an identity $\id_X$ and a modifier $f:X_2\to Y_2$,
as a modifier $\id_X\ltimes f:X\times X_2\to X\times Y_2$ 
subject to the rules in Figure~\ref{fig:lr-prod},
which form a decorated version of the rules for products. 
Symmetrically, the \emph{right semi-pure product} 
of a modifier $f:X_1\to Y_1$ and an identity $\id_X$ 
is a modifier $f\rtimes \id_X:X_1\times X\to Y_1\times X$ 
subject to the rules symmetric to those in Figure~\ref{fig:lr-prod}. 

\begin{figure}[!ht] 
\renewcommand{\arraystretch}{2.6}
$$ \begin{array}{|c|} 
\hline 
\rnlprod \dfrac {f^\modi:X_2\to Y_2} 
  {(\id_X\ltimes f)^\modi:X\times X_2\to X\times Y_2} \\  
\rnlprojone \dfrac {f^\modi:X_2\to Y_2}
  {\pi_1 \circ (\id_X\ltimes f) \eqw \pi_1} \qquad 
\rnlprojtwo \dfrac {f^\modi:X_2\to Y_2}
  {\pi_2 \circ (\id_X\ltimes f) \eqs f \circ \pi_2 } \\   
\rnlprodun \dfrac{g^\modi:X\times X_2\to Y\times Y_2 \quad 
  \pi_1 \circ g \eqw \pi_1 \quad 
  \pi_2 \circ g \eqs f \circ \pi_2 }
  {g \eqs \id_X\ltimes f} \\
\hline 
\end{array}$$
\renewcommand{\arraystretch}{1}
\caption{Rules for left semi-pure products} 
\label{fig:lr-prod} 
\end{figure}

Let us add the rules for semi-pure products to the 
decorated logic for states. 
In the decorated theory of states $\Tstate_\deco$,
let us assume that there are products 
$V_i\times V_j$ and $V_i\times \unit$ and $\unit\times V_j$ 
for all locations $i$ and~$j$. 
Then it is easy to check that the expansion of the 
decorated Equations (2)$_d$ below 
gets interpreted as Equations (2) in Remark~\ref{rem:states-eq}. 
We use the simplified notations $\id_i=\id_{V_i}$ and $\tu_i=\tu_{V_i}\,$
and $\perm_{i,j}=\perm_{V_i,V_j}$.
Equation (2.1)$_d$ has been proved in Section~\ref{subsec:liui} and 
Equation (2.6)$_d$ will be proved in Section~\ref{subsec:uiuj}.
The other equations can be proved in a similar way. 

\begin{itemize}
\item[(2.1)$_d$] 
Annihilation lookup-update.
 $ \forall\, i\in\Loc ,\;
  u_i \circ l_i \eqs \id_\unit 
  : \unit \to \unit $
\item[(2.2)$_d$] 
Interaction lookup-lookup.
 $ \forall\, i\in\Loc ,\;
  l_i\circ \tu_i \circ l_i \eqs l_i 
  : \unit \to V_i $
\item[(2.3)$_d$] 
Interaction update-update. 
 $ \forall\, i\in\Loc ,\;\; 
  u_i \circ \pi_2 \circ (u_i\rtimes\id_i) \eqs u_i \circ \pi_2 
  : V_i\times V_i \to \unit $  
\item[(2.4)$_d$] 
Interaction update-lookup. 
 $ \forall\, i\in\Loc ,\;  
  l_i\circ u_i \eqw \id_i 
  : V_i \to V_i $
\item[(2.5)$_d$] 
Commutation lookup-lookup.
 $ \forall\, i\ne j\in\Loc ,\; 
  l_j \circ \tu_i \circ l_i \eqs \perm_{j,i} \circ l_i \circ \tu_j \circ l_j 
  : \unit \to V_i\times V_j $
\item[(2.6)$_d$] 
Commutation update-update.
 $ \forall\, i\ne j\in\Loc ,\;
   u_j \circ \pi_2 \circ (u_i\rtimes \id_j) \eqs 
    u_i \circ \pi_1 \circ (\id_i \ltimes u_j) 
    : V_i\times V_j \to \unit $ 
\item[(2.7)$_d$] 
Commutation update-lookup.
 $ \forall\, i\ne j\in\Loc ,\; 
  l_j \circ u_i \eqs \pi_2 \circ (\id_i \ltimes l_j) 
    \circ (u_i \rtimes \id_j) \circ \pi_1^{-1} 
  : V_i \to V_j $ 
\end{itemize} 

\subsection{Commutation update-update} 
\label{subsec:uiuj} 

\begin{prop}
\label{prop:uiuj} 
For each locations $i\ne j\,$,
the order of storing in the locations $i$ and $j$ does not matter.
  $$ u_j^\modi \circ \pi_2^\pure \circ (u_i\rtimes \id_j)^\modi \eqs 
    u_i^\modi \circ \pi_1^\pure \circ (\id_i \ltimes u_j)^\modi 
    : V_i\times V_j \to \unit \;.$$ 
\end{prop}

\proof
In order to avoid ambiguity, in this proof 
the projections from $V_i\times\unit$ are denoted $\pi_{1,i}$ and $\pi_{2,i}$ 
and 
the projections from $\unit\times V_j$ are denoted $\pi_{1,j}$ and $\pi_{2,j}$,
while 
the projections from $V_i\times V_j$ are denoted $\pi_{1,i,j}$ and $\pi_{2,i,j}$.
It follows from Section~\ref{subsec:prod} 
that $\pi_{1,i}$ and $\pi_{2,j}$ are isomorphisms, 
while the derived rule $(E_1^\pure)$
implies that $\pi_{2,i}\eqs\tu_i$ and $\pi_{1,j}\eqs\tu_j$. 
Using the unicity property of the observational product, 
we have to prove that 
$ l_k \circ u_j \circ \pi_{2,j} \circ (u_i\rtimes \id_j) \eqw 
    l_k \circ u_i \circ \pi_{1,i} \circ (\id_i \ltimes u_j) $
		for each location $k\,$.
\begin{itemize}
\item 
When $k\neq i,j$, let us prove independently four weak equations
$(W_1)$ to $(W_4)$:   
\small
\begin{prooftree}
\AxiomC{$ (A_2)\;\; l_k \circ u_j \eqw l_k \circ \tu_j$}
\LeftLabel{\rnwsubs}
\UnaryInfC{$l_k\circ u_j\circ \pi_{2,j}\circ (u_i\rtimes \id_j) 
  \eqw l_k \circ \tu_j \circ \pi_{2,j} \circ (u_i \rtimes \id_j) \;\; (W_1)$}
\end{prooftree}
\begin{prooftree}
\AxiomC{$\vdots$}
\LeftLabel{$(E_3^\pure)$} 
\UnaryInfC{$\tu_j \circ \pi_{2,j} \eqs \pi_{1,j}$} 
     \AxiomC{$u_i$} 
     \LeftLabel{\rnrprod}  
     \UnaryInfC{$u_i \rtimes \id_j$}
  \LeftLabel{\rnssubs}
  \BinaryInfC{$\tu_j \circ \pi_{2,j} \circ (u_i \rtimes \id_j) 
      \eqs \pi_{1,j} \circ (u_i \rtimes \id_j) $}
           \AxiomC{$u_i$}
           \LeftLabel{\rnrprojone} 
           \UnaryInfC{$\pi_{1,j} \circ (u_i \rtimes \id_j) 
               \eqs u_i \circ \pi_{1,i,j}$} 
  \LeftLabel{\rnstrans}
  \BinaryInfC{$\tu_j \circ \pi_{2,j} \circ (u_i \rtimes \id_j) 
      \eqs u_i \circ \pi_{1,i,j}$}
  \LeftLabel{\rnsrepl}
  \UnaryInfC{$l_k \circ \tu_j \circ \pi_{2,j} \circ (u_i \rtimes \id_j) 
    \eqs l_k \circ u_i \circ \pi_{1,i,j}$}
  \LeftLabel{\rnsw}
  \UnaryInfC{$l_k \circ \tu_j \circ \pi_{2,j} \circ (u_i \rtimes \id_j) 
    \eqw l_k \circ u_i \circ \pi_{1,i,j} \;\; (W_2)$}
\end{prooftree}
$$
\AxiomC{$ (A_2)\;\; l_k \circ u_i \eqw l_k \circ \tu_i$}
\LeftLabel{\rnwsubs}
\UnaryInfC{$l_k \circ u_i \circ \pi_{1,i,j} \eqw l_k \circ \tu_i \circ \pi_{1,i,j}
   \;\;(W_3)$}
\DisplayProof
\qquad\qquad  
\AxiomC{$\vdots$}
\LeftLabel{$(E_3^\pure)$}
\UnaryInfC{$\tu_i \circ \pi_{1,i,j} \eqs \tu_{V_i\times V_j}$}
\LeftLabel{\rnssubs}
\UnaryInfC{$l_k \circ \tu_i \circ \pi_{1,i,j} \eqs l_k \circ \tu_{V_i\times V_j}$}
\LeftLabel{\rnsw}
\UnaryInfC{$l_k \circ \tu_i \circ \pi_{1,i,j} \eqw l_k \circ \tu_{V_i\times V_j}\;\;(W_4)$}
\DisplayProof
$$ 
\normalsize 
Equations $(W_1)$ to $(W_4)$ together with the transitivity rule for $\eqw$
give rise to the weak equation
  $ l_k\circ u_j\circ \pi_{2,j} \circ (u_i\rtimes \id_j) \eqw 
  l_k \circ \tu_{V_i\times V_j}$. 
A symmetric proof shows that 
  $ l_k\circ u_i\circ \pi_{1,i} \circ (\id_i \ltimes u_j) \eqw 
  l_k \circ \tu_{V_i\times V_j}$.
With the symmetry and transitivity rules for $\eqw$,  
this concludes the proof when $k\neq i,j$.

\item 
When $k=i$, it is easy to prove that 
$l_i\circ u_i\circ \pi_{1,i} \circ (\id_i\ltimes u_j) \eqw \pi_{1,i,j}$,
as follows.
\small
\begin{prooftree}
\AxiomC{$ (A_1)\;\; l_i \circ u_i \eqw \id_i$}
\LeftLabel{\rnwsubs}
\UnaryInfC{$l_i \circ u_i \circ \pi_{1,i} \circ (\id_i \ltimes u_j) 
  \eqw \pi_{1,i} \circ (\id_i\ltimes u_j)$}
    \AxiomC{$u_j$}
    \LeftLabel{\rnlprojone}
    \UnaryInfC{$\pi_{1,i} \circ (\id_i\ltimes u_j) \eqw \pi_{1,i,j}$}
  \LeftLabel{\rnwtrans}
  \BinaryInfC{$l_i \circ u_i \circ \pi_{1,i} \circ (\id_i \ltimes u_j) 
    \eqw \pi_{1,i,j}$}
\end{prooftree}
\normalsize
Now let us prove that 
$l_i\circ u_j\circ \pi_{2,j}\circ (u_i\rtimes \id_j) \eqw \pi_{1,i,j} $,
as follows.
\small 
\begin{prooftree}
\AxiomC{$ (A_2)\;\; l_i \circ u_j \eqw l_i \circ \tu_j$}
\LeftLabel{\rnwsubs} 
\UnaryInfC{$l_i \circ u_j \circ \pi_{2,j} \eqw l_i \circ \tu_j \circ \pi_{2,j} $}
      \AxiomC{$\vdots$}
      \LeftLabel{$(E_3^\pure)$}
      \UnaryInfC{$\tu_j \circ \pi_{2,j} \eqs \tu_{\unit \times V_j}$}
      \LeftLabel{\rnsrepl}
      \UnaryInfC{$l_i \circ \tu_j \circ \pi_{2,j} 
          \eqs l_i \circ \tu_{\unit \times V_j}$}
      \LeftLabel{\rnsw}
      \UnaryInfC{$l_i \circ \tu_j \circ \pi_{2,j} 
          \eqw l_i \circ \tu_{\unit \times V_j}$}
   \LeftLabel{\rnwtrans}
   \BinaryInfC{$l_i \circ u_j \circ \pi_{2,j} 
        \eqw l_i \circ \tu_{\unit \times V_j}$}
   \LeftLabel{\rnwsubs}
   \UnaryInfC{$l_i \circ u_j \circ \pi_{2,j} \circ (u_i \rtimes \id_j) 
     \eqw l_i \circ \tu_{\unit \times V_j} \circ (u_i \rtimes \id_j) \;\; (W'_1) $}
\end{prooftree}
\begin{prooftree} 
\AxiomC{$\vdots$}
\LeftLabel{$(E_2^\pure)$}
\UnaryInfC{$\tu_{\unit \times V_j} \eqs \pi_{1,j}$}
\LeftLabel{\rnssubs}
\UnaryInfC{$\tu_{\unit \times V_j} \circ (u_i \rtimes \id_j) 
  \eqs \pi_{1,j} \circ (u_i \rtimes \id_j)$}
    \AxiomC{$u_i$}  
    \LeftLabel{\rnrprojone}
    \UnaryInfC{$\pi_{1,j} \circ (u_i \rtimes \id_j) \eqs u_i \circ \pi_{1,i,j}$}
  \LeftLabel{\rnstrans}
  \BinaryInfC{$\tu_{\unit \times V_j} \circ (u_i \rtimes \id_j) \eqs 
     u_i \circ \pi_{1,i,j}$}
  \LeftLabel{\rnsrepl}
  \UnaryInfC{$l_i \circ \tu_{\unit \times V_j} \circ (u_i \rtimes \id_j) 
    \eqs l_i \circ u_i \circ \pi_{1,i,j}$}
  \LeftLabel{\rnsw}
  \UnaryInfC{$l_i \circ \tu_{\unit \times V_j} \circ (u_i \rtimes \id_j) 
    \eqw l_i \circ u_i \circ \pi_{1,i,j} \;\;(W'_2)$}
\end{prooftree} 
\begin{prooftree}
      \AxiomC{$ (A_1)\;\; l_i \circ u_i \eqw \id_i$}
      \LeftLabel{\rnwsubs}
      \UnaryInfC{$l_i \circ u_i \circ \pi_{1,i,j} \eqw \pi_{1,i,j} \;\;(W'_3)$}
\end{prooftree} 
\normalsize
Equations $(W'_1)$ to $(W'_3)$ and the transitivity rule for $\eqw$ 
give rise to $l_i\circ u_j\circ \pi_{2,j} \circ (u_i\rtimes \id_j) 
\eqw \pi_{1,i,j} $.
With the symmetry and transitivity rules for $\eqw$,  
this concludes the proof when $k=i$.
\item 
The proof when $k=j$ is symmetric to the proof when $k=i$.
\end{itemize}
\qed

\section*{Conclusion}

In this paper, decorated proofs are used for proving 
properties of states. To our knowkedge, such proofs are new. 
They can be expanded in order to get the usual proofs, however 
decorated proofs are more concise and closer to the syntax; 
in the expanded proof the notion of effect is lost. 
This approach can be applied to other computational effects, 
like exceptions \cite{DDFR12a,DDFR12b}. 

\bibliographystyle{eptcs}

\end{document}